\documentclass[prl,aps,twocolumn]{revtex4}
\usepackage{epsfig,amsmath}
\usepackage{bm}
\def\Fbox#1{\vskip1ex\hbox to 8.5cm{\hfil\fboxsep0.3cm\fbox{%
\parbox{8.0cm}{#1}}\hfil}\vskip1ex\noindent}
\newcommand{\B}[1]{{\bm{#1}}}
\newcommand{\C}[1]{{\mathcal{#1}}}

\newcommand{\Onecol} {\begin{widetext} \onecolumngrid} 
\newcommand{\Twocol} {\end{widetext} \twocolumngrid} 

\newcommand{\be}{\begin{equation}}
\newcommand{\ba}{\begin{array}}
\newcommand{\bea}{\begin{eqnarray}}
\newcommand{\bfi}{\begin{figure}}
\newcommand{\ee}{\end{equation}}
\newcommand{\ea}{\end{array}}
\newcommand{\eea}{\end{eqnarray}}
\newcommand{\efi}{\end{figure}}

\def\Re{${\C R}\mkern-3.1mu e$} 
\def\RE{{\C R}\mkern-3.1mu e} 
\newcommand{\Sub}[1]{_{_{\text {#1}}}} 
\begin{document}
\title{Drag Reduction by Polymers in Wall Bounded Turbulence}
\author{Victor S. L'vov, Anna Pomyalov, Itamar
Procaccia and Vasil Tiberkevich} \affiliation{ Dept. of Chemical
Physics, The Weizmann Institute of Science, Rehovot, 76100 Israel}
\begin{abstract}
We address the mechanism of drag reduction by polymers in
turbulent wall bounded flows. On the basis of the equations of
fluid mechanics we present a quantitative derivation of the
``maximum drag reduction (MDR) asymptote" which is the maximum
drag reduction attained by polymers. Based on Newtonian
information only we prove the existence of drag reduction, and
with one experimental parameter we reach a quantitative agreement
with the experimental measurements.

\end{abstract}
\maketitle

The addition of few tens of parts per million (by weight) of
long-chain polymers to turbulent fluid flows in channels or pipes
can bring about a reduction of the friction drag by up to 80\%
\cite{49Toms,75Vir,97VSW,00SW}. This phenomenon of ``drag
reduction" is well documented and is used in technological
applications from fire engines (allowing a water jet to reach high
floors) to oil pipes. In spite of a large amount of experimental
and simulational data, the fundamental mechanism has remained
under debate for a long time \cite{69Lu,90Ge,00SW}. In such
wall-bounded turbulence, the drag is caused by momentum
dissipation at the walls. For Newtonian flows (in which the
kinematic viscosity is constant) the momentum flux is dominated by
the so-called Reynolds stress, leading to a logarithmic
(von-Karman) dependence of the mean velocity on the distance from
the wall \cite{00Pope}. However, with polymers, the drag reduction
entails a change in the von-Karman log law such that a much higher
mean velocity is achieved. In particular, for high concentrations
of polymers, a regime of maximum drag reduction is attained (the
``MDR asymptote"), independent of the chemical identity of the
polymer \cite{75Vir}, see Fig. 1.
\begin{figure}
\centering
\includegraphics[width=0.5\textwidth]{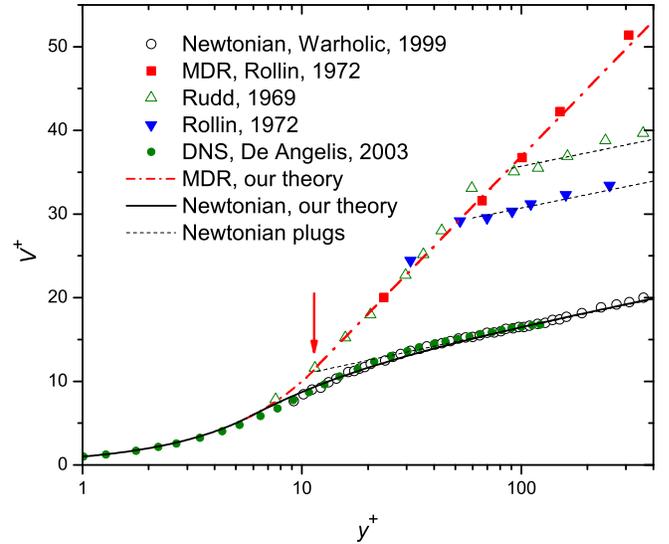}
\caption{Mean normalized velocity profiles as a function of the
normalized distance from the wall during drag reduction. The data
points from numerical simulations (green circles) \cite{03ACLPP}
and the experimental points (open circles) \cite{99WMH} represent
the Newtonian results. The black solid line is our theory
Eq.(\ref{V+}), which for large $y^+$ agrees with von-Karman's
logarithmic law of the wall (\ref{LLW}). The red data points
(squares) \cite{72RS} represent the Maximum Drag Reduction (MDR)
asymptote. The dashed red curve represents our theory for the
profile (\ref{S+VE}) which for large $y^+$ agrees with the
universal law (\ref{final}). The arrow marks the crossover from
the viscous linear law (\ref{vcs}) to the asymptotic logarithmic
law (\ref{final}). The blue filled triangles \cite{72RS} and green
open triangles \cite{69Rud} represent the cross over, for
intermediate concentrations of the polymer, from the MDR asymptote
to the Newtonian plug. Our theory is not detailed enough to
capture this cross over properly.} \label{profiles}
\end{figure}
In this Letter we elucidate the fundamental mechanism for this
phenomenon: while momentum is produced at a fixed rate by the
forcing, polymer stretching results in a suppression of the
momentum flux from the bulk to the wall. Accordingly the mean
velocity in the channel must increase. We derive a new logarithmic
law for the mean velocity with a slope that fits existing
numerical and experimental data. The law is universal, thus
explaining the MDR asymptote.

Turbulent flows in a channel are conveniently discussed
\cite{00Pope} for fixed pressure gradients $p'\equiv -\partial
p/\partial x$ where $x$, $y$ and $z$ are the lengthwise,
wall-normal and spanwise directions respectively. The length and
width of the channel are usually taken much larger than the
mid-channel height $L$, making the latter a natural re-scaling
length for the introduction of dimensionless (similarity)
variables. Thus the Reynolds number \Re, the normalized distance
from the wall $y^+$ and the normalized mean velocity $V^+(y^+)$
(which is in the $x$ direction with a dependence on $y$ only) are
defined by
\begin{equation}
\RE \equiv {L\sqrt{\mathstrut p' L}}/{\nu_0}\ , \ y^+ \equiv {y
\RE }/{L} \ , \ V^+ \equiv {V}/{\sqrt{\mathstrut p'L}} \ ,
\label{red}
\end{equation}
where $\nu_0$ is the kinematic viscosity. One of the most famous
universal aspects of Newtonian turbulent channel flows is the
``logarithmic law of the wall" which in these coordinates is
expressed as
\begin{equation}
\label{LLW} V^+(y^+) =\kappa\Sub {K}^{-1}\ln y^+ + B\,, \quad{\rm
for}~ y^+ \gtrsim 30 \,,
\end{equation}
where the von-Karman constant $\kappa\Sub K\approx 0.436 $ and
intercept $B\approx 6.13$ \cite{97ZS}. For $y^+<10$ one observes a
viscous sub-layer, $V^+(y^+)=y^+$, Fig. \ref{profiles}. The riddle
of drag reduction is then introduced in relation to this universal
law: in the presence of long chain polymers the mean velocity
profile $V^+(y^+)$ (for a fixed value of $p'$ and channel
geometry) changes dramatically. For sufficiently large
concentration of polymers $V^+(y^+)$ saturates to a new
(universal, polymer independent) ``law of the wall" \cite{75Vir},
\begin{equation}
\label{VLLW} V^+(y^+) =11.7\ln y^+ -17\,, \quad{\rm for}~ y^+
\gtrsim 12 \ .
\end{equation}
For smaller concentration of polymers the situation is as shown in
Fig. \ref{profiles}. The Newtonian law of the wall~(\ref{LLW}) is
the black solid line for $y^+\gtrsim 30$. The MDR
asymptote~(\ref{VLLW}) is the dashed red line. For intermediate
concentrations the mean velocity profile starts along the
asymptotic law~(\ref{VLLW}), and then crosses over to the so
called ``Newtonian plug" with a Newtonian logarithmic slope
identical to the inverse of von-Karman's constant. The region of
values of $y^+$ in which the asymptotic law~(\ref{VLLW}) prevails
was termed ``the elastic sub-layer" \cite{75Vir}. The relative
increase of the mean velocity (for a given $p'$) due to the
existence of the new law of the wall~(\ref{VLLW}) {\em is} the
phenomenon of drag reduction. Thus the main theoretical challenge
is to understand the origin of the new law~(\ref{VLLW}), and in
particular its universality, or independence of the polymer used.
A secondary challenge is to understand the concentration dependent
cross over back to the Newtonian plug. In this Letter we argue
that the phenomenon can be understood mainly by the influence of
the polymer stretching on the $y^+$-dependent effective viscosity.
The latter becomes a crucial agent in carrying the momentum flux
from the bulk of the channel to the walls (where the momentum is
dissipated by friction). In the Newtonian case the viscosity has a
negligible role in carrying the momentum flux; this difference
gives rise to the change of Eq. (\ref{LLW}) in favor of Eq.
(\ref{VLLW}) which we derive below.

The equations of motion of polymer solutions can be written as
\cite{87BCAH,94BE} :
\begin{equation}
{\partial \B U}/{\partial t} +\B U\cdot \B \nabla \B U=-\B \nabla
p +\B \nabla\cdot \B{{\cal T}} +\nu_0 \nabla^2 \B U \ , \label{FP}
\end{equation}
where $\B{{\cal T}}$ is the extra stress tensor that is due to the
polymer. Denoting the polymer end-to-end vector distance
(normalized by its equilibrium value) as $\B r$, the average
dimensionless extension tensor $ \B {{\cal R}}$ is ${\cal
R}_{ij}\equiv \langle r_i r_j \rangle$, and the extra stress
tensor is (with $\omega_{ij}\equiv \partial U_i/\partial x_j$),
\begin{equation}
\B{{\cal T}} =\nu_p \left(\B\omega\cdot \B {{\cal R}}+ \B {{\cal
R}}\cdot \B\omega^T - {\partial \B {{\cal R}}}/{\partial t} -\B
U\cdot \B \nabla \B {{\cal R}}\right) \ .
\end{equation}
Here $\nu_p$ (proportional to the polymer concentration) is the
polymeric contribution to the viscosity in the limit of zero
shear. In order to develop a transparent theory we propose to
ignore the fluctuations of $ \B {{\cal R}}$ compared to its mean.
In other words, we will take $ \B {{\cal R}}\approx \langle \B
{{\cal R}} \rangle $. Simplifying further the tensor structure,
assuming that ${\cal R}_{ij} ={\cal R}\delta_{ij}$ we can rewrite
Eq. (\ref{FP}) in the form
\begin{eqnarray}
&&{\partial U_i}/{\partial t} + U_j \nabla_j U_i=-\nabla_i P
+\nabla_j\nu\,
(\omega_{ij}+\omega_{ji}) \ , \label{FPnew}\\
&&\nu=\nu_0+\nu_p {{\cal R}}\ , \quad P=p+{\partial \nu}
/{\partial t} +\B U\cdot \B \nabla \nu \ ,
\end{eqnarray}
with a scalar ($y$-dependent) effective viscosity $\nu$.

Armed with the effective equation we establish the mechanism of
drag reduction following the standard strategy of Reynolds,
considering the fluid velocity $\B U(\B r)$ as a sum of its
average (over time) and a fluctuating part:
\begin{equation}
\B U(\B r,t) = \B V(y) + \B u(\B r,t) \ , \quad \B V(y) \equiv
\langle \B U(\B r,t) \rangle \ .
\end{equation}
For a channel of large length and width all the averages, and in
particular $\B V(y) \Rightarrow V(y)$, are functions of $y$ only.
The objects that enter the theory are the mean shear $S(y)$, the
Reynolds stress $W(y)$ and the kinetic energy $K(y)$; these are
defined respectively as
\[
S(y)\equiv d V(y)/d y \ , \ W (y)\equiv - \langle u_xu_y\rangle \
, \ K(y) = \langle |\B u|^2\rangle/2 .
\]
A well known exact relation \cite{00Pope} between these objects is
the point-wise balance equation for the flux of mechanical
momentum; near the wall (for $y\ll L$) it reads :
\begin{equation}
\label{MF} \nu (y)S(y) + W(y) = p'L \ .
\end{equation}
On the RHS of this equation we see the production of momentum flux
due to the pressure gradient; on the LHS we have the Reynolds
stress and the viscous contribution to the momentum flux, with the
latter being usually negligible (in Newtonian turbulence
$\nu=\nu_0$) everywhere except in the viscous boundary layer. The
$y$ dependence of the effective viscosity $\nu(y)$ in the elastic
layer will be shown to be crucial for drag reduction.

A second relation between $S(y)$, $W(y)$ and $K(y)$ is obtained
from the energy balance. The energy is created by the large scale
motions at a rate of $W(y) S(y)$. It is cascaded down the scales
by a flux of energy, and is finally dissipated at a rate
$\epsilon$, where $\epsilon = \nu \langle \omega^2_{ij}\rangle$.
We cannot calculate $\epsilon$ exactly, but we can estimate it
rather well at a point $y$ away from the wall. When viscous
effects are dominant, this term is estimated as $\nu (a/y)^2 K(y)$
(the velocity is then rather smooth, the gradient exists and can
be estimated by the typical velocity at $y$ over the distance from
the wall). Here $a$ is a constant of the order of unity. When the
Reynolds number is large, the viscous dissipation is the same as
the turbulent energy flux down the scales, which can be estimated
as $K(y)/\tau(y)$ where $\tau(y)$ is the typical eddy turn over
time at $y$. The latter is estimated as $y/b\sqrt{K(y)}$ where $b$
is another constant of the order of unity. We can thus write the
balance equation at point $y$ as
\begin{equation} \label{EB}
\nu (y)( {a}/{y} )^2 K(y) +{b\,{K^{3/2}(y)}}/y = W(y)S(y)\,,
\end{equation}
where the bigger of the two terms on the LHS should win. We note
that contrary to Eq. (\ref{MF}) which is exact, Eq.(\ref{EB}) is
not exact. We expect it however to yield good order of magnitude
estimates as is demonstrated below. Finally, we quote the
experimental fact \cite{75Vir,01PNVH} that outside the viscous
boundary layer
\begin{equation}
\label{WK} \frac{W(y)}{K(y)} = \Bigg \{ \begin{array}{ll} c\Sub
N^2\,,& \text {for Newtonian flow,}\\ & \\
c\Sub V^2\,, & \text {for viscoelastic flow.} \\
\end {array} \end{equation}
The coefficients $c\Sub{N}$ and $c\Sub{V}$ are bounded from above
by unity. (The proof is $|c|\equiv |W|/K\le 2|\langle
u_xu_y\rangle|/ \langle u_x^2+u_y^2\rangle\le 1$, because $(u_x\pm
u_y)^2\ge 0$). While $c\Sub{N}$ is known quite accurately,
$c\Sub{N}\approx 0.5$, the actual values of $c\Sub{V}$ varies
somewhat from experiment to experiment, such that the ratio
$c\Sub{V}/c\Sub{N}$ (which is all that we need to use below)
varies between 0.3 and 0.7. In our estimates we take
$c\Sub{V}/c\Sub{N}=0.5$.

We show now that Eqs. (\ref{MF} -- \ref{WK}) are sufficient for
deriving the Newtonian law (\ref{LLW}) {\em and} the viscoelastic
law (\ref{VLLW}) with equal ease. Begin with the Newtonian case.
The result of this derivation is not new - but we want to stress
that the same equations give rise to both the well known {\em and}
the new results. Substitute Eq. (\ref{WK}) in Eqs. (\ref{MF},
\ref{EB}) with $\nu(y)=\nu_0$, turning them into algebraic
equations for $K(y)$ and $S(y)$, and eventually to a first order
differential equation for $V(y)$. In the viscous sub-layer
$K(y)=0$, and the solution in re-scaled coordinates is
\begin{equation}\label{vcs}
V^+(y^+) =y^+\ , \quad y^+\le y^+_v \equiv a/c\Sub{N} \ ,
\end{equation}
Outside the viscous sub-layer ($y^+>y^+_v$) we find
\begin{subequations}\label{V+}\begin{eqnarray}
&&V^+(y^+)=\kappa\Sub K^{-1}\ln Y(y^+)+B - \Delta(y^+)\,,\\
\label{B} &&B= 2y^+_v-\kappa\Sub K^{-1} \ln\left[{e\, (1+
2\,\kappa\Sub K y^+_v)}/{4\kappa\Sub K}\right]\,,
\end{eqnarray} \end{subequations}
where $\kappa\Sub K=c\Sub{N}/b$ and we defined
\begin{eqnarray*}
Y(y^+)&=&\Big[y^++\sqrt{{y^+}^2-{y_v^+}^2+\big(2\kappa\Sub K
\big)^{-2}}\Big]\Big/ 2\,,\\
\Delta (y^+)&=&\frac{2\, \kappa\Sub K^2 {y_v^+}^2+ 4 \kappa\Sub K
\big[Y(y^+)-y^+\big]+1}{2 \kappa\Sub K ^2 y^+}\ .
\end{eqnarray*}
For $y^+\gg y^+_v$ Eqs. (\ref{V+}) turns into (\ref{LLW}) since
$Y(y^+)\to y^+ $ and $\Delta (y^+)\to 0 $.

Note that Eqs (\ref{V+}) pertain to the whole $y^+$ domain. By
taking the experimental values of $\kappa\Sub K$ and $B$ we
compute $y^+_v\approx 5.6$ to be compared with the experimental
value of 5.5$\pm 0.5$, cf. \cite{00Pope}. The resulting Newtonian
profile, Eqs. (\ref{vcs}) and (\ref{V+}), is shown in Fig. 1 as
the black solid line. The excellent agreement with the
experimental and numerical data in the entire region of $y^+$
indicates that our balance equations are sufficiently accurate and
we can proceed to the viscoelastic case.

To see how the law (\ref{VLLW}) emerges we consider Eqs.
(\ref{MF}) and (\ref{EB}) with $y$-dependent effective viscosity.
We should warn the reader that this is not fully justified --
there are terms in the full viscoelastic equations which cannot be
simplified to the form of a space dependent viscosity.
Nevertheless we propose (and justify further below) that the terms
with effective viscosity are the main terms that allow the
momentum flux to be carried in the elastic sublayer. In other
words, when the concentration of the polymer is large enough we
can neglect the second term in favor of the first in
Eq.~(\ref{MF}) and estimate $\nu(y) = Lp'/S(y)$. Substituting this
estimate in Eq.~(\ref{EB}), neglecting the second term on the LHS,
using Eq. (\ref{WK}), and finally re-scaling the variables results
in
\begin{equation}
{\partial V^+}/{\partial y^+ }= {c\Sub N}{y^+_v}/{c\Sub V}{y^+} \
. \label{good}
\end{equation}
From here follows immediately the new ``logarithmic law of the
wall"
\begin{equation}
\label{NL} V^+(y^+) = {\kappa\Sub V}^{-1} \ln y^+ +B\Sub V
\,,\quad \kappa\Sub V={c\Sub{V}}/{c\Sub N}{y^+_ v}\,,
\end{equation}
where the intercept $B\Sub V$ is still unknown (but will be
determined momentarily). The slope of the new law is {\em
independent of the polymer concentration} and is greater than the
von-Karman slope $1/\kappa\Sub K$; the ratio of the slopes is in
fact $\kappa\Sub K y^+_v c\Sub{N}/c\Sub{V}$, which is about 4.9
according to the above estimates $c\Sub{N}/c\Sub{V}\approx 0.5$
and $y^+\approx 5.6$. Comparing with the measured ratio of slopes
in Eqs. (\ref{LLW}) and (\ref{VLLW}) which is about 5.1, we
consider our estimates to be quite on the mark. As said before,
this increase in slope {\em is} the phenomenon of drag reduction.
We stress that {\em the information gained from the Newtonian data
alone is sufficient to predict drag reduction}, since $c\Sub{V}\le
1$.

To find $B\Sub V$ we match the logarithmic law (\ref{NL}) to the
viscous sublayer solution (\ref{vcs}) by the value of $V^+(y^+)$
and its derivative. First, the logarithmic law has slope 1 at
$y^+_m =\kappa\Sub V^{-1}$. Next, matching at this point the
viscous solution $V^+(y^+_m) = \kappa\Sub V ^{-1}$ to (\ref{NL})
we find $B\Sub V = \ln(e\, \kappa\Sub V )/\kappa\Sub V$. We note
that if we substitute the experimental value $\kappa\Sub K ^{-1} =
11.7$ we find $B\Sub V=-17$ in perfect agreement with Eq.
(\ref{VLLW}). We thus write the law (\ref{NL}) in its final form
(with just one constant remaining)
\begin{equation}
V^+(y^+) = \frac{1}{\kappa\Sub V}\ln\left(e\, \kappa\Sub V\,
y^+\right) \ . \label{final}
\end{equation}

Note that this universal result is obtained without reference to
any model of the polymer dynamics, and the only assumptions are
that the polymer viscosity dominates the momentum transfer in the
elastic layer and the logarithmic law~(\ref{NL}) is valid all the
way to $y^+_m$. At this point we demonstrate that these two
assumptions, which lead to the universality of (\ref{final}), are
well supported by a closer consideration of the polymer dynamics.

The fundamental reason for the appearance of a $y$-dependent
viscosity is the coil-stretch transition of the polymers under
turbulent shear. This transition occurs in any reasonable model of
the polymer-turbulence interaction, the simplest of which is of
the form \cite{77BAH}
\begin{equation}
{d{\cal R}}/{dt} = -{{\cal R}}/{\tilde \tau_p}+s(y){\cal R}\ ,
\quad s(y)\equiv \sqrt{\langle\omega_{ij}^2\rangle} \ ,
\end{equation}
where $\tilde \tau_p$ is the polymer relaxation time. In the
elastic sublayer we estimate $s(y)=g\sqrt{K(y)}/y$, with some
constant $g\approx O(1)$, and expect a coil stretch transition
when $g\sqrt{K(y)}/y =1/\tilde \tau_p $. Defining $\tau_p\equiv
g\tilde \tau_p$ we then find in the elastic sublayer (where
polymers are stretched)
\begin{equation}
K(y) = \left(y/\tau_p\right)^2 \ . \label{Ky}
\end{equation}
This equation is important since it can be solved together with
Eqs. (\ref{MF}) and (\ref{EB}) to find $S^+(y^+)$ when the
polymers are stretched. The answer is
\begin{eqnarray}
\label{S+VE} S^+(y^+)&=&\frac{1}{\kappa\Sub Vy^+}\Big\{
\Big(\frac{y^+L}{ \RE\, \ell }\Big)\\ \nonumber
&&+\sqrt{1+\big[\big(2\kappa\Sub Ky_v^+c\Sub
V\big)^2-1\big]\Big(\frac{y^+L}{ \RE\, \ell }\Big)^2}\Big\}\,, \\
\nonumber
\end{eqnarray}
where $\ell= 2\, \kappa\Sub K y^+_v c\Sub V \tau_p \sqrt{Lp'}$.
Equation (\ref{good}) is recaptured from this, more general
equation, when \Re~ is large and $\{\dots\}\to 1$. Then {\em the
identity of the polymer is lost}, giving rise to the universal
drag reduction asymptote (\ref{final}). It can be also checked
that the matching point $y^+_m$ used above indeed connects
smoothly the viscous sublayer to the asymptotic law (\ref{final})
(see arrow in Fig.~1). We can therefore conclude that our
derivation of Eq.~(\ref{final}) is fully consistent with the
polymer dynamics, and we understand how the polymer
characteristics drop out, leading to the universal law
(\ref{final}).

The mechanism of drag reduction is then the suppression of the
Reynolds stress in the elastic sublayer. The Reynolds stress is
the main agent for momentum flux in the Newtonian case, and its
suppression results in an increase of the mean mechanical momentum
(velocity) in the channel. The increase in viscosity of course
leads to increased {\em energy} dissipation, but this is
immaterial for the phenomenon. This is the main difference between
drag reduction in wall-bounded turbulence and in homogeneous
turbulence \cite{03BDGP}. We note that the theory predicts a very
low value of $K(y)$ in the elastic layer. In reality in every wall
bounded turbulent flows there will be ejecta of parcels of fluid
with high level of $K(y)$ from the Newtonian region towards the
walls. These would be picked up in measurements, and would seem to
contradict the low value of $K(y)$ predicted above. In fact there
is no real contradiction since such random ejecta do not
contribute much to the momentum flux that is so crucial to our
discussion. We also note a term in the dynamical equations for the
viscoelastic flow of the form $d\langle r_i r_j\rangle/dt$, which
cannot be written as an effective viscosity. This term vanishes
for uncorrelated rotations of individual polymeric molecules.
However it can contribute to the energy flux from the mean shear
to turbulent fluctuations in wall bounded flows due to a possible
``correlation instability", which leads to synchronized rotation
of neighboring polymers. A more detailed consideration of such
terms should provide also the cross over behavior seen in Fig.~1.

Finally we propose direct numerical simulations in channel flows
to test our approach. Instead of simulating the viscoelastic
equations, we propose to simulate the Navier-Stokes equations with
a space dependent viscosity according to the above theory. We
predict that a viscosity profile that remains constant for $0\le
y^+\le y^+\Sub V$, and then grows linearly towards the center,
should result in drag reduction in much the same way as seen in
experiments with polymers. Such simulations would render direct
support to the views offered in this Letter. \acknowledgments
\vskip 0.5 cm We acknowledge useful discussions with Rama
Govindarajan. This work was supported in part by the US-Israel
BSF, The ISF administered by the Israeli Academy of Science, the
European Commission under a TMR grant and the Minerva Foundation,
Munich, Germany.

\end{document}